\documentclass[aps,prb,twocolumn,superscriptaddress,floatfix,amsmath,amssymb]{revtex4-2}
\usepackage{multirow}
\usepackage{graphicx}
\usepackage{bm}
\usepackage{dcolumn}
\usepackage{picture}
\usepackage[colorlinks=true,linkcolor=blue,urlcolor=blue,citecolor=blue]{hyperref}
\usepackage{natbib}
\usepackage{amsmath}
\usepackage{times}
\usepackage{float}
\usepackage{color}
\usepackage{xcolor}
\usepackage{soul}
\usepackage{amssymb}
\usepackage{booktabs}
\usepackage{adjustbox}
\usepackage{threeparttable}
\usepackage[T1]{fontenc}
\begin{document}

\newcommand{\PMIO}{Pr$_2$MgIrO$_6$}

\title{Unveiling the magnetic ground states in iridate double perovskite Pr$_{2-x}$Sr$_x$MgIrO$_6$ ( $x$ = 0, 0.5) series}
% Force line breaks with \\

\author{Abhisek Bandyopadhyay}
\email[email:]{abhisek.ban2011@gmail.com}
\altaffiliation{These authors contributed equally to this work}
\affiliation{ISIS Neutron and Muon Source, STFC, Rutherford Appleton Laboratory, Chilton, Didcot, Oxon OX11 0QX, United Kingdom}
\affiliation{Department of Physics, Ramashray Baleshwar College (A Constituent Unit of Lalit Narayan Mithila University, Darbhanga), Dalsingsarai, Samastipur, Bihar 848114, India}

\author{Debu Das}
\altaffiliation{These authors contributed equally to this work}
\affiliation{School of Materials Science, Indian Association for the Cultivation of Science, Jadavpur, Kolkata-700032, India}
%\author{M. R. Lees}
%\affiliation{Department of Physics, University of Warwick, Coventry CV4 7AL, United Kingdom}

%\author{G. B. G. Stenning}
%\affiliation{ISIS Neutron and Muon Source, STFC, Rutherford Appleton Laboratory, Chilton, Didcot, Oxon OX11 0QX, United Kingdom}
%\author{M. Isobe}
%\affiliation{Max-Planck-Institut für Festkörperforschung, Heisenbergstr. 1, D-70569 Stuttgart, Germany}

%\author{Arvind K. Yogi}
%\affiliation{UGC-DAE Consortium for Scientific Research, Indore-452001, India}

%\author{D. Khalyavin}
%\affiliation{ISIS Neutron and Muon Source, STFC, Rutherford Appleton Laboratory, Harwell Science and Innovation Campus, Oxfordshire OX11 0QX, United Kingdom}
\author{Dheeraj Kumar Pandey}
\affiliation{Department of Physics, Ramashray Baleshwar College (A Constituent Unit of Lalit Narayan Mithila University, Darbhanga), Dalsingsarai, Samastipur, Bihar 848114, India}

\author{C. Ritter}
\email[email:]{ritter@ill.fr}
\affiliation{Institut Laue-Langevin, 71 Avenue des Martyrs, CS 20156, 38042 Grenoble Cedex 9, France}

\author{D. T. Adroja}
\email[email:]{devashibhai.adroja@stfc.ac.uk}
\affiliation{ISIS Neutron and Muon Source, STFC, Rutherford Appleton Laboratory, Chilton, Didcot, Oxon OX11 0QX, United Kingdom}
\affiliation{Highly Correlated Matter Research Group, Physics Department, University of Johannesburg, Auckland Park 2006, South Africa}

\author{Sugata Ray}
\email[email:]{sugataray2006@gmail.com}
\affiliation{School of Materials Science, Indian Association for the Cultivation of Science, Jadavpur, Kolkata-700032, India}

\date{\today}

\begin{abstract}
We report here the results of a detailed magnetic, thermodynamic, and neutron powder diffraction (NPD) studies carried out on the double perovskite iridates Pr$_{2-x}$Sr$_x$MgIrO$_6$ ($x = 0$ and 0.5). 
Temperature dependent bulk DC susceptibility data clearly reveals a sharp antiferromagnetic (AFM) transition at $\sim$14.5 K in \PMIO ($x$ = 0). Next, a weaker signature of an AFM transition at a lower temperature ($\sim$6 K) is observed in $x$ = 0.5 {\it i.e.,} Pr$_{1.5}$Sr$_{0.5}$MgIrO$_6$ (PSMIO1505). The observed magnetic transitions are further corroborated by the presence of anomalies around the same temperatures in our $T$-dependent specific heat results. The charge states of both Pr and Ir cations have been confirmed to be the expected ones (3+ for Pr in both the compounds, while Ir is in a pure 4+ state for $x$ = 0 and in a mixed 4+/5+ state for $x$ = 0.5) from the core-level x-ray photoemission spectroscopy (XPS) measurements. Using neutron powder diffraction (NPD) the magnetic ground states and the magnetic moment values were determined for both compounds. Both the Pr- and Ir-sites undergo AFM ordering below the respective transition temperatures, designated by the propagation vector $k$ = ($\frac{1}{2}$, 0, $\frac{1}{2}$), in both the compounds. 

\begin{description}
\item[Keyword]
Spin orbit coupling (SOC), 5$d$ iridate, Neutron Powder Diffraction, Magnetic structure, Magnetic ordering
\end{description}
\end{abstract}

\maketitle

\section{\label{sec:levelI}Introduction:}
Open-shell transition metal oxides are capable of developing a rich variety of electronic and magnetic ground states. %\textcolor{red}{SR: can you kindly check this sentence, as Clemens suggested to reformulate it?}, 
The 3$d$ transition metal oxides revealed many interesting physical properties, such as spin, charge and orbital ordering, metal-insulator transition, colossal magnetoresistance, colossal electroresistance, high-$T_C$ superconductivity, half-metallic ferro/antiferro/ferrimagnetism, quantum criticality, magnetoelectric multiferroic, etc. due to the dominant competition between only inter-electronic interaction and hopping energies \cite{Mott_1990,Hubbard_1963,fazekas1999lecture,Fisher_PRL_1988,Imada_MIT_review,Own_PRB_PFVO,TOKURA_CMR_1999}. On the other hand, the relative interplay of the relativistic interaction between an electron's spin and its orbital motion, described by the coupling term $\lambda \mathbf{L} \cdot \mathbf{S}$, crystal field, and electron correlation starts influencing the physical properties in heavier 4$d$ and 5$d$ transition metal oxide systems, giving rise to a plethora of possibilities in terms of the unconventional magnetic and electronic ground states  \cite{Kim_prl_2008,Kim_science_2009,Jackelli_prl_2009,Hasan_review_2010,Wang_prl_2011,Kim_prl_2012,Wan_prb_2011,Agrestini_prb_2018,Nag_prb_2018}.
\par
5$d$ Iridium-based oxides have been serving as a potential paradigm for investigating a diverse variety of quantum mechanical ground states, starting from the spin-orbit entangled $J_{eff}$ = $\frac{1}{2}$ state \cite{Kim_prl_2008,Kim_science_2009} to topological superconductivity \cite{Kim_science_2014,Kim_nature_2015,Zhao_nature_2016} to the quantum spin liquid (QSL) phase \cite{Jackelli_prl_2009,Krempa_review_2014,Rau_review_2016}. On top of it, magnetism in these Ir-oxide systems has turned out to be an exciting branch due to the possibility of multiple Ir-valences (3+, 4+, 5+, 6+) and the resulting different Ir-$d$ electron configurations, which inevitably brings changes in the effective SOC strength, covalency, hybridization, hopping, noncubic crystal field, bandwidth, and magnetic exchange interactions in a real solid depending on the Ir-valence/$d$-electron occupancy, leading to the variety of novel magnetic ground states \cite{Nag_prb_2018,Nag_prb1_2018,Kayser_Ir6+_2014,TKD_prb_2014,BHOWAL2020166827,Swarup_prb_Irchargestate_2025,Own_prb_2022,Own_prm_2024}. For example, in case of Ir$^{4+}$ (5$d^5$) oxide systems, strong SOC introduces splitting of the crystal-electric-field-split low-spin triply degenerate Ir-$t_{2g}$ orbitals into filled $J_{eff}$ = $\frac{3}{2}$ quartet and half-filled $J_{eff}$ = $\frac{1}{2}$ doublet bands, and then, upon inclusion of moderate electronic correlations, the $J_{eff}$ = $\frac{1}{2}$ band is further split into narrower subbands (completely filled lower Hubbard and an empty upper Hubbard bands with a finite energy gap in between) and forms spin-orbital-entangled local moment $J_{eff} = \frac{1}{2}$ pseudo-spin state \cite{Kim_prl_2008,Kim_science_2009}. It would be relevant to highlight in this context that Guo {\it et al.}
\cite{Clemens_PRB_2016}, for the very first time among the pyrochlore iridate family, made the direct experimental manifestation of the ordered Ir$^{4+}$ magnetic moments in Nd$_2$Ir$_2$O$_7$ using NPD within the framework of a spin-orbit-coupled $J_{eff} = \frac{1}{2}$ state. %Importantly, this $J_{eff}$ = $\frac{1}{2}$ state has been the topic of intense recent condensed matter and quantum materials research activities in search for unique quantum phases \cite{Wang_prl_2011,Chaloupka_prl_2010}.
On the other hand, the pentavalent (Ir$^{5+}$) 5$d^4$ iridates are predicted to possess nonmagnetic $J_{eff}$ = 0 singlet ground state in the strong SOC limit. Several experimental reports however do not support such a $J_{eff}$ = 0 state, rather, clearly demonstrate the development of intrinsic finite magnetism together with a disordered magnetic ground state in real materials \cite{Nag_prb_2018,Own_prb_2022,Nag_prb1_2018,Nag_prl_2016,Own_jpcm_2024}. The 5$d^3$ hexavalent (Ir$^{6+}$) iridate counterparts, on the contrary, hold merely negligible SOC effect together with stronger Ir-O covalency, giving rise to the low ordered spin-only moments with complex magnetic ordering \cite{Kayser_Ir6+_2014,BHOWAL2020166827,Swarup_prb_Irchargestate_2025}.
\par
At this point, the idea to explore the magnetic properties of an iridate series through a systematic tuning of the Ir-$d$ electron occupancy turns out to be quite obvious although has not been studied enough. %\textcolor{red}{SR: Can you kindly check and reformulate the next sentence, as Clemens found it as "strange"?} 
%Despite being considered such an intriguing matter of fact, surprisingly however, microscopic magnetic investigations are thus far lacking to shed lights on the true nature of the magnetic ground states for any materials class where systematic tuning over a range of Ir-$d$ electron counts is studied \cite{Phelan_prb_2015,Own_prb_2019}. 
Phelan {\it et al}, by means of bulk DC susceptibility measurements, reported the evolution of the Ir magnetic moment from 5$d^5$ Ir$^{4+}$
to 5$d^4$ Ir$^{5+}$, and a concomitant transition from the spin-orbit-driven strongly magnetic $J_{eff} = \frac{1}{2}$ to a nonmagnetic $J_{eff}$ = 0 Ir state in the La$_{11-x}$Sr$_x$Ir$_4$O$_{24}$ ($x$ = 1, 3, 5) series with nearly mutually disconnected Ir-O octahedra \cite{Phelan_prb_2015}. Later, we worked on the Sr-doped iridate double perovskite (DP) Pr$_{2-x}$Sr$_x$MgIrO$_6$ ($x$ = 0, 0.5, 1) series \cite{Own_prb_2019} and showed using combined structural, electronic, and bulk magnetometry characterizations a smooth gradual crossover from an antiferromagnetic transition for $x$ = 0 (Pr$_2$MgIrO$_6$, Ir$^{4+}$: 5$d^5$) to vanishing  magnetic ordering down to 2 K for $x$ = 1 (PrSrMgIrO$_6$, Ir$^{5+}$: 5$d^4$) via weakening of the transition in $x$ = 0.5 (Pr$_{1.5}$Sr$_{0.5}$MgIrO$_6$, mixed Ir$^{4+}$/Ir$^{5+}$).
\par
In this work, we have extended our study to members of this series Pr$_2$MgIrO$_6$ ($x$=0) and Pr$_{1.5}$Sr$_{0.5}$MgIrO$_6$ ($x = 0.5$) by detailed magnetic, thermodynamic, and microscopic neutron powder diffraction (NPD) investigations to get deeper insights into the nature of the magnetic ground states as well as of the ground state spin structures. In conjunction with the state-of-art DC susceptibility, magnetization, and specific heat results, our detailed microscopic neutron powder diffraction (NPD) measurements down to 2 K unveil the long-range AFM ordering of both the Pr and Ir sublattices within the magnetic propagation vector $k$ = ($\frac{1}{2}$, 0, $\frac{1}{2}$) below the respective transition temperatures for either of the compounds. Further, our NPD data analyses demonstrate that the Pr-site magnetic moment is confined within the $a-b$ plane for both the compounds. While, the Ir-site has a large component of the ordered magnetic moment along the unit cell $c$-axis direction for the undoped compound, the $c$-axis direction for the Sr-doped system, on the contrary, doesn't appear to be the dominant component of the Ir magnetic moment. Sr-doping eventually reduces the Pr$^{3+}$-Pr$^{3+}$ as well Ir$^{4+}$-Ir$^{4+}$ exchange coupling strengths through the inclusion of random Pr-Sr and Ir$^{4+}$-Ir$^{5+}$/Ir$^{5+}$-Ir$^{5+}$ connectivity, resulting in the weakening of the magnetic ordering transition as well as the reduction in the ordered Pr and Ir magnetic moments for the Sr-doped system Pr$_{1.5}$Sr$_{0.5}$MgIrO$_6$ relative to the parent \PMIO \ compound.     

\section{\label{sec:levelII}EXPERIMENTAL SECTION:}
Polycrystalline Pr$_{2-x}$Sr$_x$MgIrO$_6$ ($x$ = 0, 0.5; abbreviated as PMIO and PSMIO1505, respectively) samples were synthesized using the conventional solid-state reaction technique reported in ref. \cite{Own_prb_2019}. 
The phase-purity of both the samples was examined through x-ray powder diffraction (XRPD) using a Bruker AXS D8 Advance X-ray diffractometer. The preliminary structural characterizations were performed on the collected XRPD data using the Rietveld refinement method within the \textsc{FullProf} software package \cite{Carvajal_1993}. The chemical homogeneity and cation-stoichiometry of either of the samples were verified in a JEOL JSM-7500F field-emission scanning electron microscope with an Oxford Instruments energy-dispersive X-ray (EDX) spectrometer. The accurate cation-stoichiometry of both the compounds was additionally confirmed by inductively coupled plasma-optical emission spectroscopy (ICP-OES) using a Perkin Elmer Optima 2100 DV instrument. Core-level x-ray photoemission spectroscopy (XPS) measurements were carried out at room temperature using an Omicron electron spectrometer, equipped with a Scienta Omicron Sphera analyzer and Al $K\alpha$ monochromatic source with an energy resolution of 0.5~$eV$. The temperature dependent DC magnetic susceptibility of both the materials was measured in a Magnetic Property Measurement System (MPMS, Quantum Design) equipped with a vibrating sample magnetometer in the $T$-range of 2–300~K and in magnetic fields up to $\pm$7~T. The specific heat was measured as a function of temperature in both zero and applied high magnetic fields (up to 70 kOe) using a Physical Property Measurement System (PPMS, Quantum Design). Neutron powder diffraction data were recorded at room temperature on powder samples of about 4 gm of Pr$_2$MgIrO$_6$ and Pr$_{1.5}$Sr$_{0.5}$MgIrO$_6$ using the high resolution powder diffractometer D2B at the Institut Laue Langevin (ILL), Grenoble, France, with $\lambda$ = 1.594 {\AA}. In addition, high intensity neutron powder diffraction data were collected on the diffractometer D1B, as well at the ILL, with $\lambda$ = 2.524 {\AA}. Data were accumulated for 2 hours at 20 K (above $T_N$) and at 2 K (below $T_N$) for \PMIO with an additional temperature scan with $\Delta$$T$ = 0.5 K between these two temperatures taking a diffraction data for every 5 minutes. For Pr$_{1.5}$Sr$_{0.5}$MgIrO$_6$, the NPD data were measured for 19 hours at 20 K and at 2 K. All measurements were done using a double wall vanadium container to reduce the strong absorption of Ir. Rietveld refinement of the data were performed using the program \textsc{FullProf} \cite{Carvajal_1993}. Magnetic symmetry analysis was performed using the programs \textsc{Basireps} \cite{Clemens_2011,IR_MSG_NPD,Carvajal_NPD_Magnetic_2025}, \textsc{MAXMAGN} \cite{Perez_2015}, and \textsc{ISODISTORT} \cite{campbell_2006}.

\section{\label{sec:levelII}RESULTS AND DISCUSSION:}
\subsubsection{Structural characterizations by x-ray and neutron powder diffractions}
\begin{figure*}
\centering
\includegraphics[width=1\textwidth]{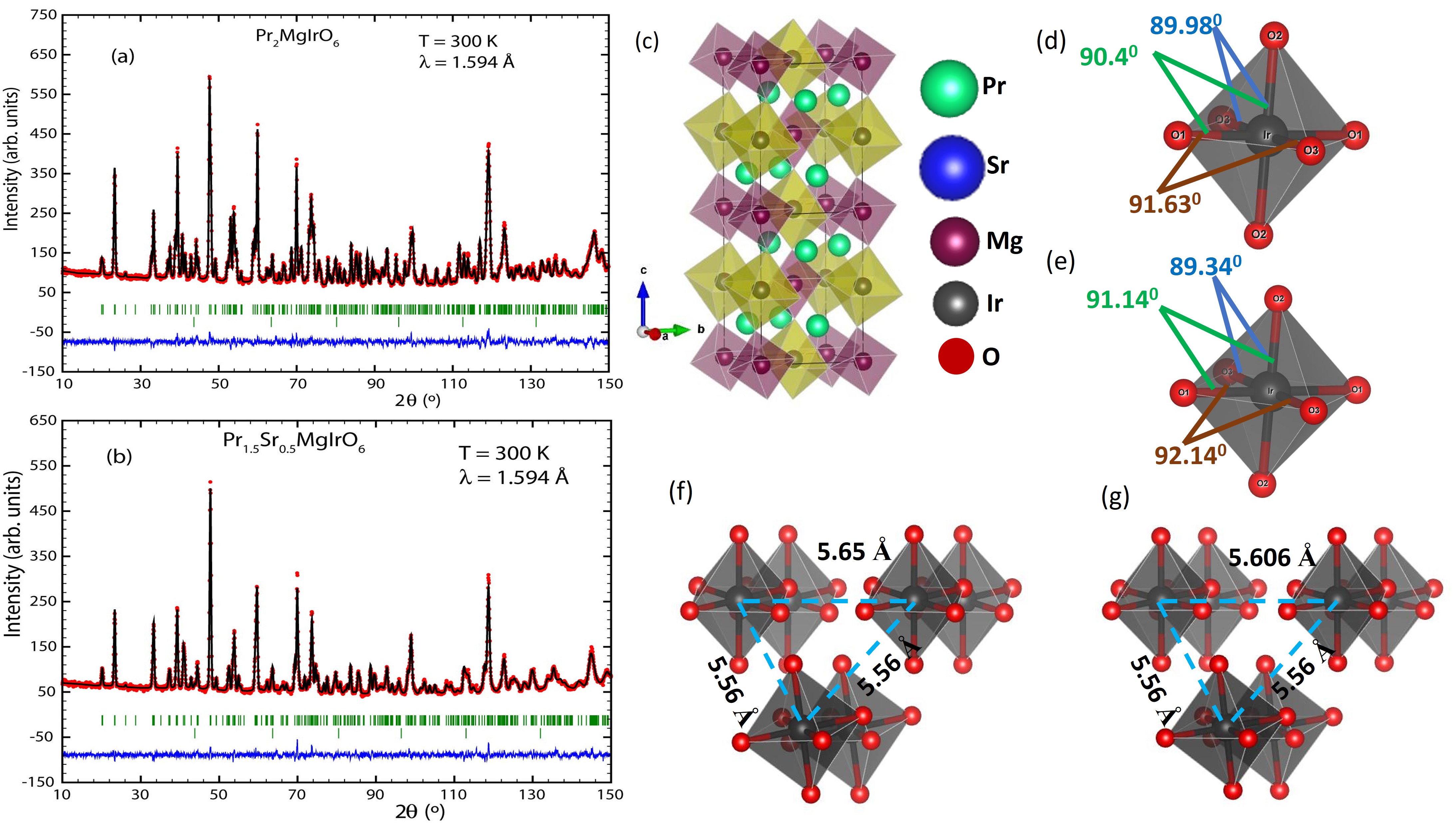}
\caption{ Rietveld refined NPD of (a) \PMIO and (b) {Pr$_{1.5}$Sr$_{0.5}$MgIrO$_6$}. Open black circles represent the experimental data and the continuous red line represents the calculated pattern. The blue line represents the difference between the observed and calculated patterns while the vertical green tick marks indicate the positions of the Bragg reflections of the compounds in $P$2$_{1/n}$ (upper row) and of the vanadium sample holder (lower row). (c) show the crystal structure of \PMIO. % and {Pr$_{1.5}$Sr$_{0.5}$MgIrO$_6$} respectively.
The rotational distortions (change in O-Ir-O bond angles) within the IrO$_6$ octahedral unit for \PMIO~ (d), and {Pr$_{1.5}$Sr$_{0.5}$MgIrO$_6$} (e) samples. In addition, the
extent of the geometric frustration within Ir-triangular units is shown for the \PMIO (f),~and {Pr$_{1.5}$Sr$_{0.5}$MgIrO$_6$} (g) compounds.}
\label{FIG:NPD}
\end{figure*}

The phase purity and crystal structures of both $x$ = 0 and 0.5 samples of the Pr$_{2-x}$Sr$_x$MgIrO$_6$ series were examined by the Rietveld refined room temperature x-ray powder diffraction (XRPD) patterns. 
%As displayed in Fig.~\ref{FIG:A1},
Both the compounds crystallize in pure single-phase with monoclinic $P2_1/n$ crystal symmetry, which is consistent with our previous work \cite{Own_prb_2019}. 
%\begin{figure}[h]
%\centering
%\includegraphics[width=1\linewidth]{Fig1_XRD.jpg}
%\caption{ Rietveld refined XRPD of (a) \PMIO~and (b) {Pr$_{1.5}$Sr$_{0.5}$MgIrO$_6$}. Open black circles represent the experimental data and the continuous red line represents the calculated pattern. The blue line represents the difference between the observed and calculated patterns while the vertical green tick marks indicate the positions of the Bragg reflections.}
%\label{FIG:A1}
%\end{figure}
Following the XRPD results, the high resolution neutron powder diffraction (NPD) patterns further confirm that both the compounds crystallize in the double perovskite structure with general formula $A_2$$BB^{\prime}$O$_6$ within space group $P$2$_{1/n}$ as reported before by Mugavero et al.~\cite{Mugavero_JSSC}, and Bandyopadhyay et al.~\cite{Own_prb_2019}. Both samples are single phase (in agreement with XRPD results) with a slight Mg/Ir cation site-disorder on the $B$ sites of about 6\% in Pr$_2$MgIrO$_6$ and about 2.5\% in Pr$_{1.5}$Sr$_{0.5}$MgIrO$_6$. Fig.~\ref{FIG:NPD} (a) \& (b) display the Rietveld refinements of the room temperature NPD data of both compounds. The refined lattice parameters, atomic positions, and site-occupancies are shown in Table~\ref{tab:refinement}.

%\end{center}

Clearly, due to large ionic size mismatch between Sr$^{2+}$ (1.44 {\AA}) and Pr$^{3+}$ (1.12 {\AA}), upon Sr-doping, the Pr/Sr layered ordering at the $A$-site influences the Mg/Ir cation ordering at the $B$-site \cite{Own_prb_2019,Own_prb_2017}, resulting in a more $B$-site ordered scenario by means of smaller Mg-Ir site-disorder in the Sr-doped sample relative to the undoped one. Comparing the structural details of the two compounds as determined from the NPD data one can see that the $x$ = 0.5 compound possesses a higher extent of IrO$_6$ octahedral rotational distortions (in terms of deviation of O-Ir-O bond angles from ideal 90$^{\circ}$, shown in Fig. \ref{FIG:NPD} (d) and (e)) compared to the $x$ = 0 compound in order to accommodate the bigger sized Sr$^{2+}$ cations within the cubooctahedral cavity in the DP structure. This results in slightly larger noncubic crystal fields around the Ir-O octahedra of the doped compound. In addition, interestingly, as depicted in Fig. \ref{FIG:NPD} (f), (g), the Sr-doping reduces the difference in the two types of Ir-Ir edge-lengths of the iridium triangular units, thus exposing the Pr$_{1.5}$Sr$_{0.5}$MgIrO$_6$ system to greater geometric frustration than the Pr$_2$MgIrO$_6$.      

\begin{table}[t]
\centering

\caption{Refined structural parameters and fit goodness factors obtained from the NPD data refinements of both the compounds. The cation disorder on the $B/B^{\prime}$ sites was refined keeping site occupations fixed to the nominal values.
(a) Pr$_2$MgIrO$_6$ (300 K): $a=5.507(1)$\,\AA, $b=5.653(2)$\,\AA, $c=7.835(8)$\,\AA; 
$\alpha=\gamma=90^\circ$, $\beta=89.964^\circ$; $R_B=3.3$, and $\chi^2=1.7$.
(b) Pr$_{1.5}$Sr$_{0.5}$MgIrO$_6$ (300 K): $a=5.541(5)$\,\AA, $b=5.606(7)$\,\AA, 
$c=7.855(2)$\,\AA; $\alpha=\gamma=90^\circ$, $\beta=90.041^\circ$; 
$R_B=3.9$,  and $\chi^2=1.5$.}
\label{tab:refinement}

\begin{tabular}{l l c c c c c}
\toprule
Sample & Atoms & Occupancy & $x$ & $y$ & $z$ \\
% & $B$ ($\times 10^3$ \AA$^2$) 
\midrule
\multirow{6}{*}{Pr$_2$MgIrO$_6$} 
 & Pr   & 1.0   & 0.4888(7) & 0.0528(4) & 0.249(1)   \\
 & Mg1  & 0.94 & 0         & 0         & 0          \\
 & Ir1  & 0.06 & 0         & 0         & 0           \\
 & Ir2  & 0.94 & 0.5       & 0.5       & 0           \\
 & Mg2  & 0.06 & 0.5       & 0.5       & 0           \\
 & O1   & 1.0   & 0.202(1) & 0.3016(9) & 0.0465(8)   \\
 & O2   & 1.0   & 0.5900(4) & 0.4743(4) & 0.2464(8)   \\
 & O3   & 1.0   & 0.298(1) & 0.7922(9) & 0.0490(7) &   \\
\midrule
\multirow{6}{*}{Pr$_{1.5}$Sr$_{0.5}$MgIrO$_6$}
& Pr   & 0.75  & 0.493(1) & 0.0409(4) & 0.246(1)   \\
& Sr   & 0.25  &0.493(1) & 0.0409(4) & 0.246(1)   \\
 & Mg1  & 0.975 & 0         & 0         & 0           \\
 & Ir1  & 0.025 & 0         & 0         & 0          \\
 & Ir2  & 0.975 & 0.5       & 0.5       & 0           \\
 & Mg2  & 0.025 & 0.5       & 0.5       & 0         \\
 & O1   & 1.0   & 0.209(1) & 0.299(1) & 0.0403(1)  \\
 & O2   & 1.0   & 0.5786(6) & 0.4811(6) & 0.252(1)   \\
 & O3   & 1.0   & 0.294(1) & 0.785(1) & 0.045(1)   \\
\bottomrule
\end{tabular} 
\end{table}

\subsubsection{Core-level X-ray photoemission spectroscopy (XPS)}
\begin{figure}[h]
\centering
\includegraphics[width=1\linewidth]{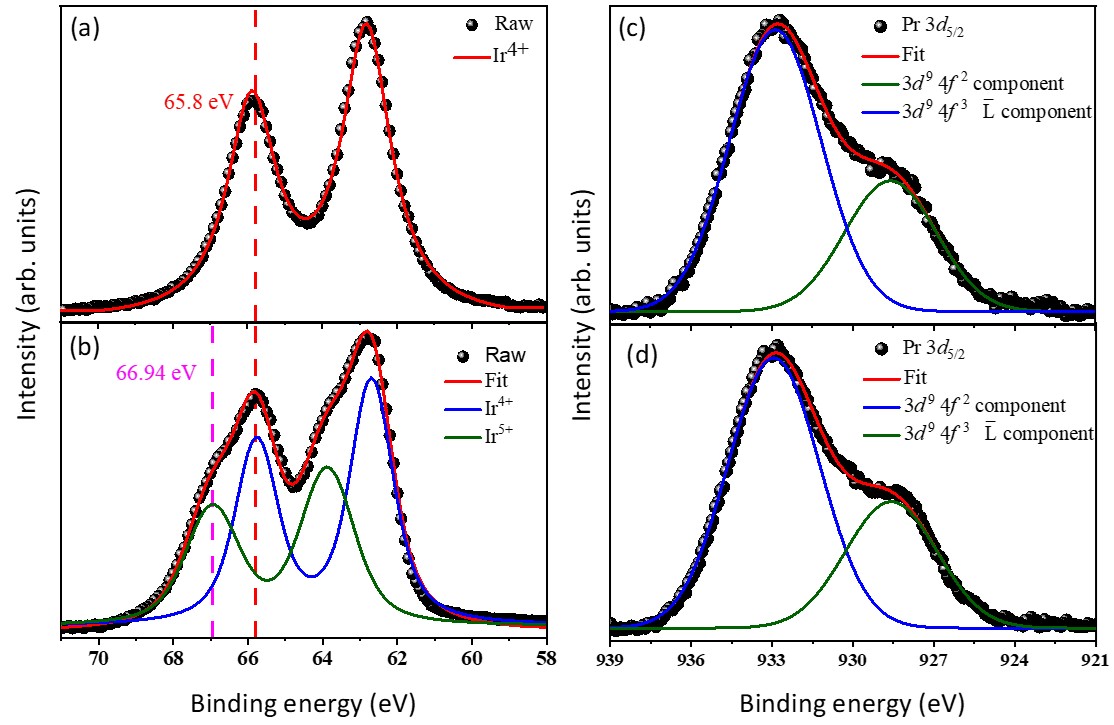}
\caption{\label{fig:frog} Ir 4$f$ core level XPS spectra
(shaded black circles) along with the fitting (red solid line) for (a) Pr$_{2}$MgIrO$_6$ and (b) Pr$_{1.5}$Sr$_{0.5}$MgIrO$_6$ samples.~Pr 3$d_{5/2}$ core level XPS spectra (shaded black circles) along with the respective fittings (red solid line) for the (c) Pr$_{2}$MgIrO$_6$ and (d) Pr$_{1.5}$Sr$_{0.5}$MgIrO$_6$ samples.}
\label{FIG:XPS}
\end{figure}

Considering the fact that both Pr and Ir can take multiple oxidation states (e.g. Pr$^{3+}$, Pr$^{4+}$ and Ir$^{3+}$, Ir$^{4+}$, Ir$^{5+}$, Ir$^{6+}$) in a given material and that these valences critically influence the respective magnetic ground state properties, it is of utmost importance to determine the Pr and Ir valence states in both PMIO and PSMIO1505 compounds. Consequently, the experimental Ir-4$f$ and Pr-3$d_{5/2}$ core-level XPS spectra were collected and shown in the left and right panels, respectively, of Fig. \ref{FIG:XPS} along with the theoretical fittings. For \PMIO, the peaks of the Ir-4$f$ core-level XPS spectrum display nearly symmetric shape, and therefore, have been well captured using a single spin-orbit split doublet. As displayed in Fig.~\ref{FIG:XPS} (a), the positions of the higher (4$f_{5/2}$) and lower (4$f_{7/2}$) binding energy (BE) features, the spin-orbit energy separation, and the ratio of the respective peak intensities, refer to the pure 4+ charge state of Ir in \PMIO. On the other hand, the Ir-4$f$ core-level XPS spectra of Pr$_{1.5}$Sr$_{0.5}$MgIrO$_6$ (see Fig. \ref{FIG:XPS} (b)) appears to be strongly asymmetric, and hence, has been accounted for by a combination of two spin-orbit doublets. We assign the lower BE doublet (blue doublet) to the Ir$^{4+}$ charge state, while the higher BE doublet (green doublet) to the Ir$^{5+}$ charge state. This gives rise to the fraction of the Ir$^{4+}$ charge with respect to the Ir$^{5+}$ species, estimated from the ratio of the respective integrated intensity [{\it i.e.}, ratio of the area under the blue doublet of Ir$^{4+}$ to that of green doublet from Ir$^{5+}$, shown in Fig. \ref{FIG:XPS} (b)], to be $\approx$ 1, which is in complete agreement with the stoichiometrically expected 4.5+ valence of Ir in the doped compound Pr$_{1.5}$Sr$_{0.5}$MgIrO$_6$. Further, the Pr-3$d_{5/2}$ core-level XPS spectra were collected and analyzed for both the $x$ = 0 \& 0.5 samples, the results of which are shown in Fig. \ref{FIG:XPS} (c) and (d). Instead of a clean singlet, pronounced lower BE shoulder appears for both the samples. The observed double peak features of the Pr-3$d_{5/2}$ photoemission spectra are due to final state configurations from 3$d^9$ 4$f^3$ $\bar{L}$ ($\bar{L}$ denotes ligand hole on the O 2$p$ orbital) component on the lower BE side (shoulder) and from the 3$d^9$ 4$f^2$ component on the higher BE side (Main peak) \cite{Ogasawara_Pr_prb,Guzik_Pr_2014}. Importantly, the peak shape, energy positions, and the relative peak intensities of the Pr 3$d_{5/2}$ line are very much similar among the two measured samples, and match exactly well with those of reported in Pr$_2$O$_3$ and PrAlO$_3$ systems \cite{Ogasawara_Pr_prb,Guzik_Pr_2014}, thus confirming pure 3+ valence of Pr in both the $x$ = 0 and 0.5 compounds.
 
%Importantly, XPS confirms that the Pr ions retain their trivalent state (Pr$^{3+}$). 
By affirming the Pr and Ir-valence states to their respective stoichiometrically expected values, the fulfillment to charge-neutrality of both the samples is perfectly achieved. This further indirectly supports the absence of any detectable oxygen vacancy and/or elemental off-stoichiometry in any of these two studied systems.

\subsubsection{DC magnetic susceptibility and DC magnetization}
\begin{figure*}
\centering
\includegraphics[width=1\textwidth]{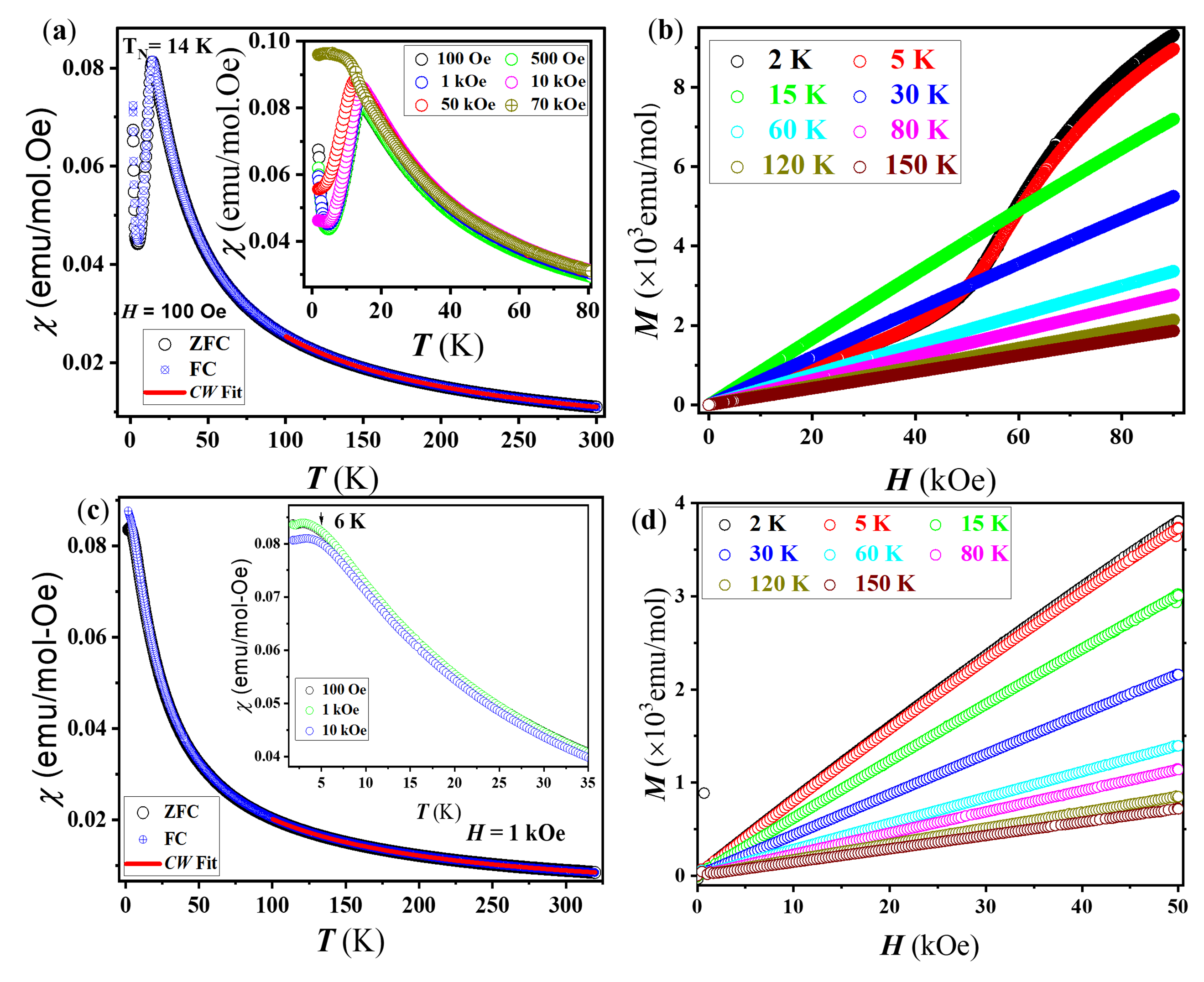}
\caption{ Zero-field-cooled (open black circles) and field-cooled (crossed blue circles) DC magnetic susceptibilities as a function of temperature for (a) \PMIO~and (c) Pr$_{1.5}$Sr$_{0.5}$MgIrO$_6$; Inset of (a) reveals the ZFC DC susceptibility curves at various applied magnetic fields, pointing to a field-induced change in the nature of the magnetic transition. Isothermal magnetization ($M-H$) curves at different temperatures for (b) \PMIO~ and (d) Pr$_{1.5}$Sr$_{0.5}$MgIrO$_6$.}
\label{FIG:MT}
\end{figure*}

Following the structural and electronic characterizations, temperature dependent both zero-field-cooled (ZFC) and field-cooled (FC) DC magnetic susceptibility measurements ($\chi$ vs. $T$) were carried out on both the samples in different applied  magnetic fields. As depicted in the main panel of Fig.~\ref{FIG:MT} (a), the 1 kOe $\chi$-$T$ data from the undoped sample \PMIO~ reveal a sharp antiferromagnetic (AFM) transition at around 14.5~K ($T_N$), which is consistent with the previous work \cite{Own_prb_2019}. %\textcolor{red} {SR: Can you please check the last sentence in the context of citation, as Clemens feel that "we are mentioning too often our previous PRB paper on this PrIr series, and therefore, the referees could question the inclusion of the magnetic measurements into this article as they are not new".} 
Additionally, we have performed $\chi(T)$ measurements between 2 and 300 K at several high magnetic fields (10, 50, and 70 kOe), the results of which are displayed in the inset of Fig.~\ref{FIG:MT} (a). The AFM transition remains unaltered in applied fields up to $H$ = 50 kOe, while the 14 K peak disappears at $H$ = 70 kOe and nearly flattened $\chi(T)$ curves are evident below the transition temperature. This could possibly indicate a field-driven spin-reorientation transition phenomenon in \PMIO, which is further corroborated by the emergence of a field-induced metamagnetic transition at applied $H>50$ kOe in the 2 and 5 K $M-H$ isotherms as shown in Fig. \ref{FIG:MT} (b). Such a field-induced change in the low-$T$ magnetic response implies that at $H >$ 50 kOe, the system turns into a polarized ferromagnet through a gradual suppression of the AFM phase. The featured metamagnetic transition ({\it i.e.}, field-induced change in spin orientation) in materials, particularly, with strong spin-orbit coupling and lattice distortions, is of great interest for spintronic applications, such as magnetic switching, memory devices, and logic elements.

For the Sr-doped Pr$_{1.5}$Sr$_{0.5}$MgIrO$_6$ sample, the AFM transition is significantly suppressed and moved to the lower temperature of $\sim$6~K, resulting probably from the weakening of the magnetic exchange interactions due to the reduced density of Ir$^{4+}$ ions as a result of inclusion of Ir$^{5+}$ species upon Sr$^{2+}$-doping, causing increased effective spatial separation between the Ir$^{4+}$ ions in the Sr-doped compound \cite{Own_prb_2019}. The relatively higher extent of geometrical
frustration within the isosceles Ir-triangular network [see
Fig. \ref{FIG:NPD} (g), (h)] of the doped compound compared to the undoped one possibly impedes the AFM transition in the Sr-doped case.

Well above the respective AFM transition temperatures, the 1 kOe field-cooled DC susceptibility data of both the $x$ = 0 and 0.5 samples were fitted using the Curie–Weiss (CW) law, $\chi(T) = \chi_0 + \frac{C}{T -\theta_{\mathrm{W}}}$, where, $\chi_0$ being the temperature-independent paramagnetic susceptibility, $C$ and $\Theta_W$ be the Curie constant and Weiss temperature, respectively. As shown in the main panel of Fig. \ref{FIG:MT} (a), (c), the fitting results yield $\mu_{eff}$ to be $\sim$ 5.15 and 4.54 $\mu_B$/f.u. for the $x$ = 0 and 0.5 samples, respectively, while, the Weiss temperature, $\Theta_W$, to be $\sim$ -36.68 K for \PMIO~ and $\sim$ -35.89 K for Pr$_{1.5}$Sr$_{0.5}$MgIrO$_6$, reflecting predominant nearest-neighbor AFM exchange. %It is to be noted that the estimated values of the CW fitting parameters are in perfect agreement with the previous work \cite{Own_prb_2019}\textcolor{red} {used ref where samples were different}. 
Notably, despite having very similar AFM exchange interactions (in terms of negative $\Theta_W$ values) in either of the compounds, the $x = 0.5$ sample reveals lower AFM transition temperature and consequently, produces larger frustration index, $f$ (= $\frac{\Theta_W}{T_N}$ $\sim 6$), than the undoped \PMIO~ material ($f \sim 2.5$), which is consistent with the relative extent of geometrical frustration within the Ir-triangular units of the respective samples. % parameters match with the previously reported values and confirm the idea that Pr does not possess any kind of magnetic coupling with the magnetic B site. 

\subsubsection{Heat capacity ($C_p$)}
\begin{figure*}
\centering
\includegraphics[width=1\textwidth]{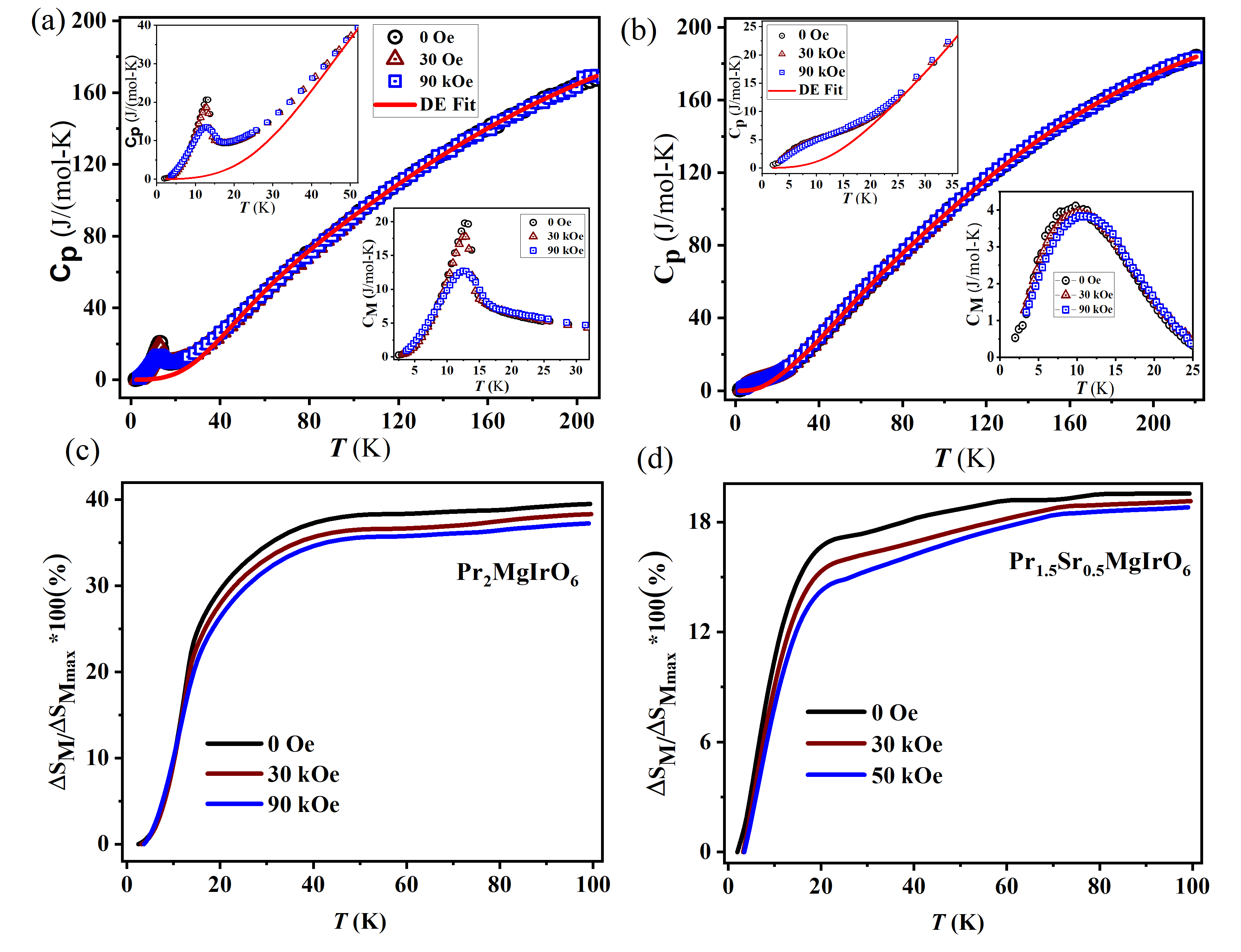}
\caption{ Specific heat of (a) \PMIO~ and (b) Pr$_{1.5}$Sr$_{0.5}$MgIrO$_6$ at different applied magnetic field (Inset: Magnetic contribution to the specific heat from Debye-Einstein fitting). (c) and (d) depict the corresponding entropy change in \PMIO~ and Pr$_{1.5}$Sr$_{0.5}$MgIrO$_6$ respectively.}
\label{FIG:HC}
\end{figure*}

In order to probe the thermodynamic signature of magnetic ordering, heat capacity $C_p$ measurements were performed on a small piece of pressed pellet of \PMIO~and Pr$_{1.5}$Sr$_{0.5}$MgIrO$_6$ as a function of temperature under zero and different applied magnetic fields between 1.8 and 300 K, the results of which are shown in the main panels of Fig.~\ref{FIG:HC} (a) and (b). The respective upper insets display the expanded views of the $C_p-T$ data in the low-$T$ regions from 25 K to 1.8 K for \PMIO~ and 15 K to 1.8 K for the Pr$_{1.5}$Sr$_{0.5}$MgIrO$_6$ sample. The zero-field $C_p(T)$ data of \PMIO~ reveal a sharp $\lambda$-like anomaly at around 14 K, supporting the transition to a three-dimensional long-range magnetic order, which is in agreement with our $T$-dependent DC susceptibility results shown in Fig. \ref{FIG:MT}(a). For the Sr-doped material PSMIO1505, a weak but clear anomaly is evident at around 6 K in the zero-field $C_p-T$ data (top left inset to Fig. \ref{FIG:HC}(b)), which is again consistent with the measured bulk susceptibility data, referring to a possible magnetic transition. In the absence of a suitable nonmagnetic analogue, to further investigate the magnetic contribution of the $C_p$ data of both the systems, the lattice part was extracted by modeling the respective high-temperature $C_p$ data (well above the magnetic transition to eliminate any effect of short-range magnetic correlations from the estimated lattice contribution) using the Debye-Einstein equation with a combination of one Debye and three Einstein terms contributing to the acoustic and optical phonon modes, respectively. The resulting fit yields a Debye temperature of $\Theta_D \approx$ 276 K for \PMIO~ and 156 K for the Sr-doped compound, while the Einstein temperatures turn out to be $\Theta_{E_1} =$ 700 K, $\Theta_{E_2} =$ 692 K, and $\Theta_{E_3} =$ 506 K for the undoped system and $\Theta_{E_1} =$ 251 K, $\Theta_{E_2} =$ 246 K, and $\Theta_{E_3} =$ 600 K for the 25\% Sr-doped compound. We have verified that the sum of the coefficients of the Debye and Einstein terms correspond to 10 vibrational modes for either of the compounds, which exactly matches with the total number of atoms per formula unit in both the $x = 0$ and $0.5$ cases, thus validating our lattice part fitting. This high-$T$ fitted data is then extrapolated down to the lowest 1.8 K and considered as $C_L$ (see insets to Fig. \ref{FIG:HC} (a) and (b)). After subtracting the lattice part from the measured total $C_p$ data, we get the correlated magnetic contributions to the heat capacity, {\it i.e.}, the magnetic specific heat $C_M$. As displayed in the bottom right insets of Fig.~\ref{FIG:HC} (a) and (b), the $T$-dependence of magnetic specific heat $C_M$ of both the $x$ = 0 and 0.5 compounds clearly reveals nearly field-independent anomalies at the onset of their respective magnetic transitions, supporting further the long-range magnetic ordering in either of these two systems. For \PMIO, although the magnetic transition temperature does not shift with applied $H$, the sharp $\lambda$-like peak feature present at low applied magnetic fields gets suppressed and evolves into a broad anomaly in applied $H > 50$ kOe, indicating a field-induced polarization into a ferromagnetic state, which is consistent with the results from our $T$-dependent DC susceptibility and field-dependent isothermal magnetization measurements [see Fig.~\ref{FIG:MT} (a), (b)]. On the other hand, the field-independent character of the observed broad feature around the magnetic transition temperature in the $C_M-T$ data of Pr$_{1.5}$Sr$_{0.5}$MgIrO$_6$ (see bottom right inset to Fig. \ref{FIG:HC} (b)) cannot be attributed to a two-level Schottky anomaly effect, but possibly relates to  a magnetic ordering transition. %\textcolor{red}{SR: Clemens suggested to reformulate the previous sentence for making it more clearly understandable; could you please have a look and revise it?} 
At the same time the absence of a sharp $\lambda$-anomaly even in the zero-field $C_M-T$ data reflects weakening of the magnetic ordering transition in the Sr-doped sample relative to the parent \PMIO.  
\begin{figure}
\centering
\includegraphics[width=0.5\textwidth]{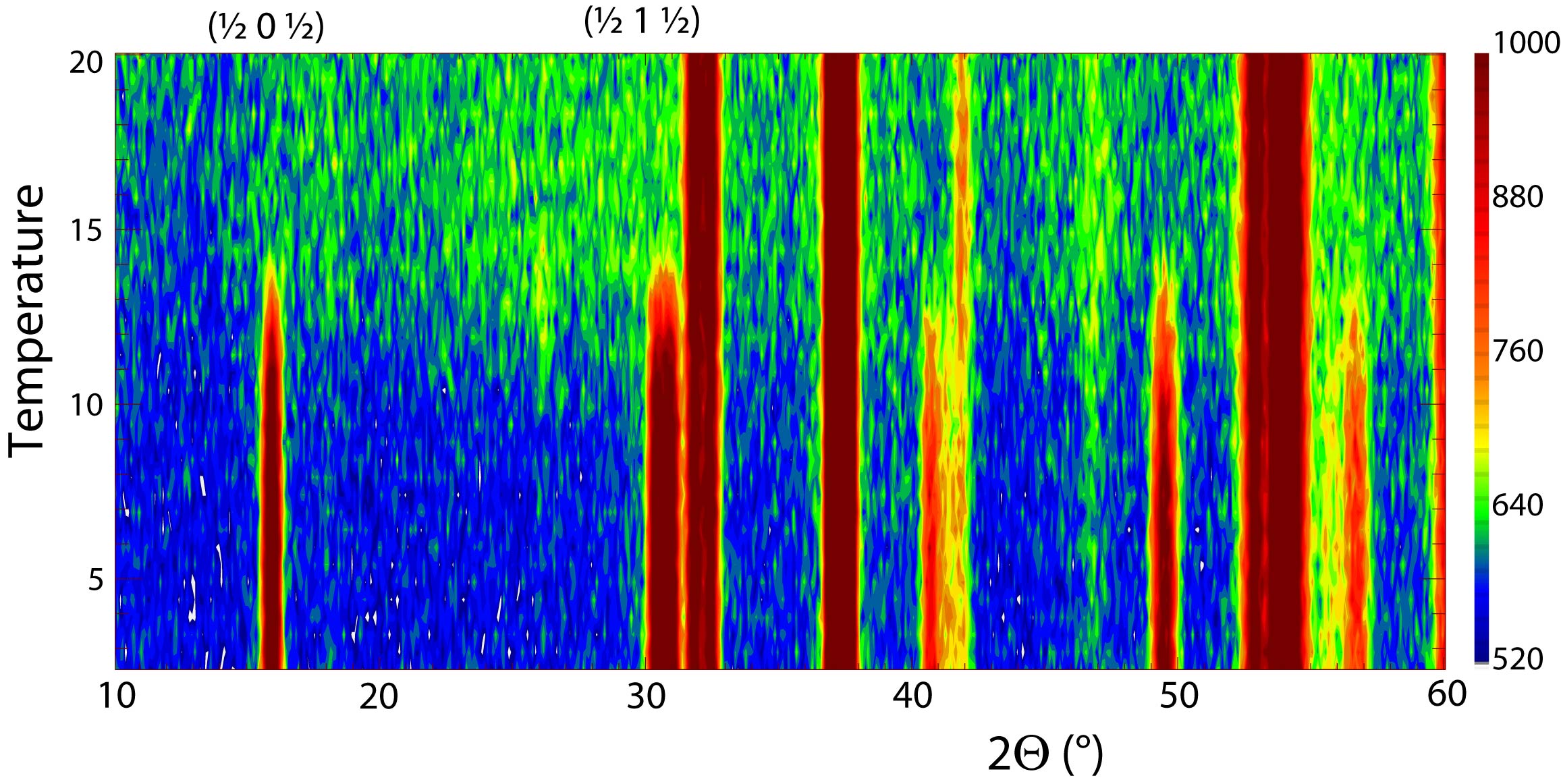}
\caption{Low angle part of the thermodiffractogram of \PMIO~showing the appearance of magnetic Bragg peaks indexed with $k$ = ($\frac{1}{2}$, 0, $\frac{1}{2}$). The scale of the color-coded intensity contour map is given at the right for reference.}
\label{FIG:TDG}
\end{figure}
\subsubsection{Magnetic Entropy}
To further elucidate the nature of the magnetic phase transition in both the \PMIO~ and Pr$_{1.5}$Sr$_{0.5}$MgIrO$_6$ compounds, the change of magnetic entropy ($\Delta S_M$) has been estimated using the relation, $\Delta S_M(T)$ = $\int\frac{C_M}{T}dT$, where $C_M$ is the magnetic contribution to the heat capacity, shown in the insets of Fig. \ref{FIG:HC} (a) and (b). As displayed in Fig. \ref{FIG:HC} (c), the temperature dependence of the magnetic entropy change of \PMIO~ exhibits a smooth continuous evolution around the magnetic transition, ensuring the second order nature of the magnetic phase transition. It is to be noted that the expected maximum release in magnetic entropy associated with complete magnetic ordering of \PMIO and Pr$_{1.5}$Sr$_{0.5}$MgIrO$_6$ are, respectively, [$2\times R\ln 9$ ($\mathrm{Pr^{3+}}$ contribution with $J_{eff} = 4$) + $R\ln 2$ ($\mathrm{Ir^{4+}}$ contribution with $S_{eff} = 1/2$)] $\approx$ 42.30 J mol$^{-1}$K$^2$ and [$1.5\times R\ln 9$ ($\mathrm{Pr^{3+}}$ contribution with $J_{eff} = 4$) + $0.5\times R\ln 2$ ($\mathrm{Ir^{4+}}$ contribution with $S_{eff} = 1/2$) + $0.5\times R\ln 3$ ($\mathrm{Ir^{5+}}$ contribution with $S_{eff} = 1$)] $\approx$ 34.84 J mol$^{-1}$K$^2$.
The estimated amount of recovered magnetic entropy turns out to be $\sim$ 40\% and $\sim$20\%, respectively, for \PMIO~ and Pr$_{1.5}$Sr$_{0.5}$MgIrO$_6$, of the respective maximum entropy, suggesting presence of short-range magnetic correlations above the magnetic transition, as typically reckoned in frustrated magnetic systems. Such a reduced release of entropy could be due to the complex interplay between spin-orbit entanglement, crystal-electric-field effect, and incomplete magnetic ordering \cite{Matsuoka_Pr_2007, Silva_Pr_1978}. Moreover, the much smaller release of magnetic entropy in case of the Sr-doped material than the undoped compound is in consistent with the pertinent higher degree of frustration in Pr$_{1.5}$Sr$_{0.5}$MgIrO$_6$ compared to the \PMIO~case, which likely agrees with the estimated frustration index (inferred from our bulk DC susceptibility data) of these materials. 

\subsubsection{Neutron powder diffraction (NPD)}
In order to have a microscopic understanding of the true magnetic ground state as well to investigate the ground state spin structure, the neutron powder diffraction (NPD) measurements were carried out for both the \PMIO~ and Pr$_{1.5}$Sr$_{0.5}$MgIrO$_6$ materials at temperatures above and below their respective magnetic transitions. 

\begin{figure*}
\centering
\includegraphics[width=0.9\textwidth]{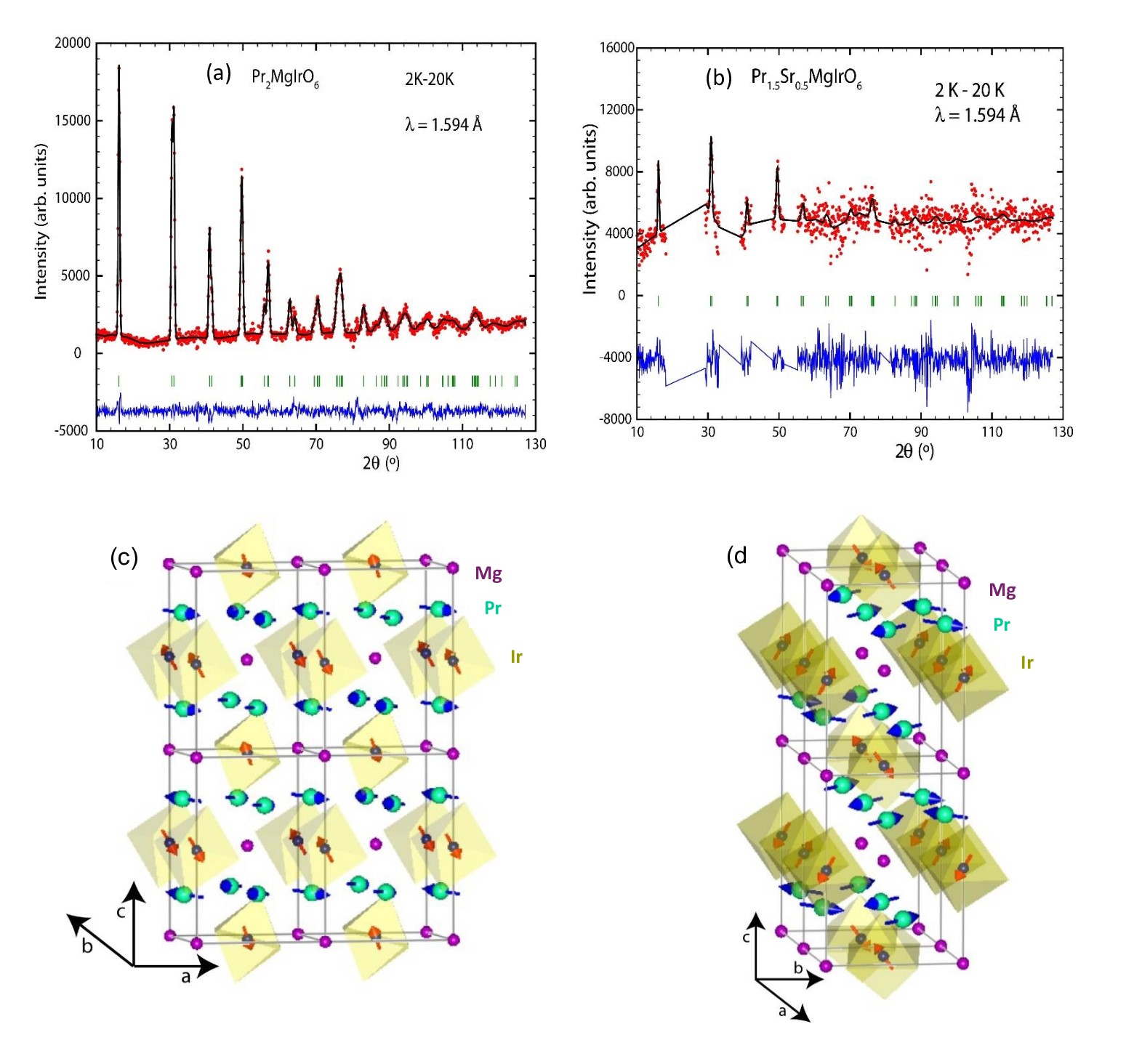}
\caption{Refinement of the purely magnetic diffraction intensity of \PMIO~(a) and Pr$_{1.5}$Sr$_{0.5}$MgIrO$_6$ (b) created by subtracting data taken at 20 K in the paramagnetic state from the data taken at base temperature of 2 K. A constant background was added to have all positive values. The closed circles (red) and the solid black line correspond to the experimental data and the calculated intensities. The blue line represents their difference and the green tick marks indicate the position of the magnetic Bragg reflections created by the magnetic propagation vector $k = (\frac{1}{2}, 0, \frac{1}{2})$. For Pr$_{1.5}$Sr$_{0.5}$MgIrO$_6$, regions with strong remaining features resulting from the subtraction of the very strong nuclear peaks were excluded from the refinements. (c) and (d): Two different views of the magnetic structure of \PMIO, where the size of the magnetic moment of Ir (red spin) has been rescaled by 50\% for easy identification. While the magnetic moment of Pr (blue spin) is mostly confined to the Pr-layer within the $a-b$-plane that of Ir has a large component along the direction of the unit cell $c$-axis.} 
\label{FIG:NPDMS}
\end{figure*}

{\it Magnetic structure of }\PMIO:

Fig. \ref{FIG:TDG} displays the thermodiffractogram of \PMIO~ between 20 and 2 K. Additional strong Bragg peaks having all the same temperature dependence appear below about 14.5 K. This temperature agrees with the magnetic transition as determined by Bandyopadhyay {\it et al.} \cite{Own_prb_2019} and also inferred from the bulk DC susceptibility and heat capacity data in our present work. The additional reflections can therefore be identified as being of magnetic origin. Fig. \ref{FIG:NPDMS} shows the difference between the data taken with high statistics at base temperature (2 K) and above the magnetic transition temperature at 20 K corresponding to the pure magnetic diffraction intensity of \PMIO. All the magnetic reflections can be indexed with the magnetic propagation vector $k = (\frac{1}{2}, 0, \frac{1}{2})$. 

Magnetic symmetry analysis with $k = (\frac{1}{2}, 0, \frac{1}{2})$ was done for the Pr and the Ir positions on the Wyckoff sites 4$e$ and 2$c$ of $P2_1/n$, respectively, using the program BASIREPS \cite{Clemens_2011}. Table \ref{Table:IR1} lists the basis vectors of the four allowed irreducible representations (IR) for the two sites along the three unit cell directions, which correspond to the four maximal magnetic spacegroups \cite{Perez_2015} allowing non zero magnetic moments on at least one of the two sites. All four IR were tested against the experimental data and Table \ref{Table:IR1} shows that IR2 gives the by far best reliability factors. For these refinements of the difference data the results of the 20 K refinement within the paramagnetic state were used. Lattice parameters, atomic coordinates, peak-shape parameters, zero-shift, and most importantly, the scale-factor of the 20 K refinement were fixed and only the magnetic moment components and the background parameters were allowed to vary in the refinement of the difference data.
The resulting magnetic structure is shown in Fig. \ref{FIG:NPDMS}, with total ordered magnetic moments amounting to $\mu_{\mathrm{Pr}}$ = 2.05(1) $\mu_B$ and $\mu_{\mathrm{Ir}}$ = 0.88(4) $\mu_B$, while, the magnetic space group is UNI  $P2_1/c.1'_a$ [$P2_1/c$] (BNS 14.80 $P_a2_1/c$) \cite{Perez_2015,campbell_2006}. Compared to the free ion Pr$^{3+}$ moment (3.58 $\mu_B$), the reduced value of ordered Pr$^{3+}$ magnetic moment (2.05 $\mu_B$) could be attributed to the emergence of crystal-electric-field (CEF)-split $J_{eff}$ state (other than the $J$ = 4 ground state of a free Pr$^{3+}$ species) of Pr$^{3+}$ influencing strongly the low-$T$ correlated Pr$^{3+}$ magnetic response \cite{Bu_PRB_2022}, and hence, the ordered magnetic moment in our \PMIO material. Notably, this ordered moment value of Pr$^{3+}$ in our present case is consistent with the results of the previously reported other Pr-based compounds \cite{Pr_moment_2017, Pr_moment_Pr2LiRuO6}.

%For PMIO ($x = 0$), a clear enhancement in magnetic scattering intensity was observed at low temperatures (e.g., 2~K), indicating the development of long-range antiferromagnetic (AFM) order. The magnetic Bragg peaks were indexed and successfully refined using the Rietveld method, revealing a commensurate AFM structure. In contrast, for the Sr-doped sample PSMIO1505 ($x = 0.5$), the magnetic signal was significantly weaker, consistent with the suppression of magnetic ordering due to the mixed-valent state of Ir. Nonetheless, subtle magnetic features could still be detected at low temperatures, suggesting the presence of short-range or weakly ordered AFM correlations. These results demonstrate how Sr doping not only reduces the magnetic transition temperature but also weakens the robustness of the long-range magnetic structure.

\begin{table*}[t]
\centering
\caption{Basis vectors (BV) of the four allowed irreducible representations (IR) for 
$\mathbf{k} = (\tfrac{1}{2}, 0, \tfrac{1}{2})$ and Wyckoff positions $4e$ (Pr) and $2c$ (Ir) of space group $P2_1/n$. 
The three BV are along the unit cell directions: BV1 = 100, BV2 = 010, BV3 = 001.}
\begin{adjustbox}{width=\textwidth}
\begin{tabular}{l ccc ccc ccc ccc c}
\toprule
 & \multicolumn{3}{c}{IR1} & \multicolumn{3}{c}{IR2} & \multicolumn{3}{c}{IR3} & \multicolumn{3}{c}{IR4} & \multirow{2}{*}{$\mu$ ($\mu_B$)} \\
\cmidrule(lr){2-4} \cmidrule(lr){5-7} \cmidrule(lr){8-10} \cmidrule(lr){11-13}
 & BV1 & BV2 & BV3 & BV1 & BV2 & BV3 & BV1 & BV2 & BV3 & BV1 & BV2 & BV3 & \\
\midrule
Pr; $x,y,z$        & + & + & + & + & + & + & + & + & + & + & + & + & \\
$-x+\tfrac{1}{2}, y+\tfrac{1}{2}, -z+\tfrac{1}{2}$ & + & - & + & + & - & + & - & + & - & - & + & - & \\
$-x,-y,-z$          & + & + & + & - & - & - & + & + & + & - & - & - & \\
$x+\tfrac{1}{2}, -y+\tfrac{1}{2}, z+\tfrac{1}{2}$ & + & - & + & - & + & - & - & + & - & + & - & + & \\
                &   &   &   & 1.25(1) & 1.62(1) & 0.15(2) &   &   &   &   &   &   & 2.05(1) \\
\midrule
Ir; $x,y,z$        & / & / & / & + & + & + & / & / & / & + & + & + & \\
$-x+\tfrac{1}{2}, y+\tfrac{1}{2}, -z+\tfrac{1}{2}$ & / & / & / & + & - & + & / & / & / & - & + & - & \\
                &   &   &   & -0.31(4) & -0.47(2) & 0.68(4) &   &   &   &   &   &   & 0.88(4) \\
\midrule
$R_{\text{Mag}}$ & \multicolumn{3}{c}{28} & \multicolumn{3}{c}{7.1} & \multicolumn{3}{c}{29} & \multicolumn{3}{c}{18.3} & \\

\bottomrule
\end{tabular}
\end{adjustbox}
\label{Table:IR1}
\end{table*}

The above mentioned identical temperature dependence of all magnetic reflections points to the fact that the magnetic ordering of both the Pr-site and the Ir-site appears simultaneously at $T_N$ (=14.5 K). While the magnetic moment of Pr-site is mostly confined to the Pr-layer within the $a-b$-plane, the magnetic moment of Ir has a large component in direction of the unit cell $c$-direction. In this context one can point to the differing pathways governing the magnetic exchange interactions. The super-super-exchange interactions Ir-O-Mg-O-Ir determining the magnetic coupling between the Ir sites are nearly isotropic as the three different Ir-O-Mg angles ($\angle$IrO1Mg = 150.3$^{\circ}$, $\angle$IrO3Mg = 149.7$^{\circ}$ within the $a-b$-plane and $\angle$IrO2Mg = 150.5$^{\circ}$ between the layers) and distances (Ir-O1-Mg = 2.017 {\AA}, 2.066 {\AA}, Ir-O2-Mg = 2.027 {\AA}, 2.024 {\AA}, Ir-O3-Mg = 2.025 {\AA}, 2.064 {\AA}) are very similar. At the same time the super-exchange interactions via Pr-O-Pr pathways between the Pr sites are highly anisotropic between intralayer and interlayer sites. There are three Pr-O-Pr intralayer interactions ($\angle$PrO1Pr = 99.6$^{\circ}$, Pr-O1 = 2.653 {\AA}, 2.404 {\AA}; $\angle$PrO2Pr = 114.0$^{\circ}$, Pr-O2 = 2.444 {\AA}, 2.359 {\AA} and $\angle$PrO3Pr = 101.4$^{\circ}$, Pr-O3 = 2.386 {\AA}, 2.607 {\AA}) embracing the three different oxygen sites and four interlayer Pr-O-Pr interactions (2 x $\angle$PrO1Pr = 102.0$^{\circ}$, Pr-O1 = 2.721 {\AA}, 2.404 {\AA} and 2 x $\angle$PrO1Pr = 158.4$^{\circ}$, Pr-O1 = 2.721 {\AA}, 2.653 {\AA}) passing solely through the O1 site and showing different Pr-O-Pr angles and Pr-O distances.

{\it Magnetic structure of} Pr$_{1.5}$Sr$_{0.5}$MgIrO$_6$:

\begin{table}[t]
\centering
\caption{Refined values of the basis vectors (BV) of IR2, total magnetic moment values $\mu$ and magnetic R-factors of the refinement of the difference data 2 K – 20 K of Pr$_{1.5}$Sr$_{0.5}$MgIrO$_6$. Listed are the values of the refinement including all 6 BVs to vary freely and others where the BVs of the Ir-site are totally or partly constrained to zero.}

\begin{adjustbox}{width=0.95\columnwidth}
\begin{tabular}{c l ccc c c}
\toprule
Model &  & \multicolumn{3}{c}{IR2} & \multirow{2}{*}{$\mu$ ($\mu_B$)} & \multirow{2}{*}{$R_{\text{Mag}}$} \\
\cmidrule(lr){3-5}
 &  & BV1 & BV2 & BV3 &  &  \\
\midrule
\multirow{3}{*}{1} 
 & Pr & 0.21(6) & 0.29(4) & -0.08(6) & 0.37(5) &  \\
 & Ir & 0.10(13) & -0.17(7) & 0.07(7) & 0.21(9) &  \\
 &    &   &   &   &   & 12.7\% \\
\midrule
\multirow{3}{*}{2} 
 & Pr & 0.28(4) & 0.23(4) & -0.06(5) & 0.37(4) &  \\
 & Ir & / & / & / & / &  \\
 &    &   &   &   &   & 17.8\% \\
\midrule
\multirow{3}{*}{3} 
 & Pr & 0.21(4) & 0.30(3) & -0.07(4) & 0.37(3) &  \\
 & Ir & / & -0.11(3) & 0.07(6) & 0.13(4) &  \\
 &    &   &   &   &   & 14.0\% \\
\midrule
\multirow{3}{*}{4} 
 & Pr & 0.19(4) & 0.29(3) & -0.07(5) & 0.35(3) &  \\
 & Ir & / & -0.13(3) & / & 0.13(3) &  \\
 &    &   &   &   &   & 14.8\% \\
\bottomrule
\end{tabular}
\end{adjustbox}
\label{Table:IR2}
\end{table}
On the other hand, the magnetic ordering temperature of the Pr$_{1.5}$Sr$_{0.5}$MgIrO$_6$ compound has been determined to be $T_N$ $\sim$ 6 K from our bulk magnetic and thermodynamic measurements (see Section III. (3) and (4)). The neutron powder diffraction data taken for this compound show that in comparison to the undoped \PMIO~ sample the intensity of the magnetic diffraction are strongly reduced. It was therefore not possible to follow its evolution with temperature and only very long measurements of 20 hours at 2 K and 20 K, and using the difference between these two datasets allowed the detection of very small magnetic Bragg peaks. These reflections appear at similar angular positions as in the undoped \PMIO~and can again be indexed with the magnetic propagation vector $k = (\frac{1}{2}, 0, \frac{1}{2})$. The same solutions, proposed from symmetry analysis as shown in Table II, are therefore valid and were tested against the data. Like in \PMIO, the magnetic structure corresponding to IR2 is again the solution for the Sr-doped compound with the best goodness of fit values, $R_{Mag}$ = 12.7\%; the refinement of the data is shown in Fig. \ref{FIG:NPDMS} (b).

Allowing all the BVs to vary freely in the refinement, the determined values of the BVs of the Ir-site are partly smaller or similar to the size of the error bars making the refinement unstable and the total magnetic moment values are not well defined (Table \ref{Table:IR2}, model 1). However, any attempt to ignore the presence of the small magnetic moment on the Ir-site leads to worsening the refinements (Table \ref{Table:IR2}, model 2). Two further refinements were therefore performed keeping BV1 of Ir (Table \ref{Table:IR2}, model 3) or BV1 and BV3 of Ir (Table \ref{Table:IR2}, model 4) constrained to zero. This allowed to stabilize the fit and to get better estimates of the total magnetic moments without reducing significantly the goodness of the refinements. Compared to the undoped compound the ordered magnetic moment values are strongly reduced in the doped system Pr$_{1.5}$Sr$_{0.5}$MgIrO$_6$ to $\mu_{\mathrm{Pr}}$
 $\sim$ 0.36(3) $\mu_B$ and $\mu_{\mathrm{Ir}}$ $\sim$ 0.13(3) $\mu_B$. Such a much reduced ordered magnetic moment of Pr$^{3+}$ in the doped system relative to the undoped one is possibly due to the fact that the bigger sized Sr-cation incorporation at the Pr-site of the doped compound evidently disrupts the local Pr-site symmetry and hence, naturally brings modification in the Pr$^{3+}$-CEF-effect relative to the undoped \PMIO, thereby, causing difference in the Pr$^{3+}$-CEF-driven low-energy $J_{eff}$ states and resulting in a reduction of the low-$T$ ordered Pr$^{3+}$ moments in this Sr-doped material with respect to the \PMIO. Another perspective could be posed here in this context that Sr-doping dilutes the strongly magnetic Ir$^{4+}$ network through the creation of weakly magnetic Ir$^{5+}$ species, which in turn softens the internal molecular field at the Pr-sublattice. Such a weakening of internal field in the Pr-site of the Sr-doped compound essentially leads to the reduced exchange-induced mixing of the low-lying Pr$^{3+}$-excited CEF states, giving much smaller ordered Pr$^{3+}$ moment in the Sr-doped sample than \PMIO. The orientation of the magnetic moments of the Pr-site in Pr$_{1.5}$Sr$_{0.5}$MgIrO$_6$ is very similar to the situation in Pr$_2$MgIrO$_6$ being again confined nearly exclusively to the Pr-layers. Due to the very small values and the large error bars, it is not possible to ascertain precisely the orientation of the moments on the Ir-site in Pr$_{1.5}$Sr$_{0.5}$MgIrO$_6$; it seems, however, the Ir magnetic moment component in direction of the unit cell $c$-axis is no longer the dominant component.

In fact, the effect of Sr-doping on the magnetic ground state behaviors of the iridate double perovskite Pr$_{2-x}$Sr$_x$MgIrO$_6$ series is of two-fold. Firstly, upon substitution of Pr with Sr in the doped system, the Pr-O-Pr exchange interaction pathways get disrupted through Sr incorporation, and consequently, the Pr-Pr exchange coupling strengths get significantly diluted as a result of reduced numbers of Pr-O-Pr connectivity via development of Pr-O-Sr/Sr-O-Sr links due to Sr-doping. Secondly, Sr-doping introduces the Ir$^{5+}$ species into the doped system along with the Ir$^{4+}$ charges of the parent \PMIO. Considering Ir$^{4+}$ to be strongly magnetic \cite{Kim_prl_2008,Cao_PRB_2007,Cao_PRB_2013} and Ir$^{5+}$ weakly magnetic \cite{Nag_prl_2016,Nag_prb_2018,Nag_prb1_2018,Own_prb_2022,Own_jpcm_2024} and/or ideally nonmagnetic in the strong spin-orbit coupling limit \cite{Khaliullin_PRL_2013,Chen_PRB_2011}, the Sr-doping-induced creation of Ir$^{5+}$ ions eventually increases the spatial separation between the magnetic Ir$^{4+}$ species due to the enhanced (reduced) density of Ir$^{5+}$ (Ir$^{4+}$) ions upon hole doping, which results in the suppression of magnetic exchange interaction strengths between the Ir$^{4+}$ magnetic moments in the Sr-doped compound relative to the undoped \PMIO through the formation of Ir$^{4+}$-O-Mg-O-Ir$^{5+}$/Ir$^{5+}$-O-Mg-O-Ir$^{5+}$ exchange linkages over the primary Ir$^{4+}$-O-Mg-O-Ir$^{4+}$ pathways. While the first factor certainly weakens the AFM ordering transition among the Pr-sites, the second aspect is at the origin of undermining the AFM ordering between the Ir-sites in the doped compound.      

\section{CONCLUSIONS}
The magnetic ground state properties and the magnetic structures of the rare-earth based iridium double perovskite Pr$_{2-x}$Sr$_x$MgIrO$_6$ ($x = 0$, 0.5) series of compounds have been determined using magnetic, thermodynamic and neutron powder diffraction methods. XPS data confirmed that the oxidation state of Pr remains Pr$^{3+}$ on doping while the valence of Ir changes from Ir$^{4+}$ in the parent compound to mixed Ir$^{4+}$/Ir$^{5+}$ in the x=0.5 compound. The non-doped \PMIO undergoes a well-defined antiferromagnetic transition at about 14.5 K below which the Pr- and the Ir-sublattices order simultaneously with the magnetic propagation vector $k = (\frac{1}{2}, 0, \frac{1}{2})$ and $\mu_{Pr} = 2.05(1) \mu_{B}$ and $\mu_{Ir} = 0.88(4) \mu_{B}$, magnetic space group UNI $P2_1/c.1'_a $[$P2_1/c$]. Doping with Sr leads to a weakening of the AFM interactions, a decrease of $T_N$ to about 6 K and a strong reduction of the magnetic moment values to $\mu_{Pr} = 0.36(3) \mu_{B}$ and $\mu_{Ir} = 0.13(3) \mu_{B}$ while notwithstanding the same magnetic structure is kept. The decrease of the AFM interactions in Pr$_{1.5}$Sr$_{0.5}$MgIrO$_6$ can be attributed to the emergence of a mixed Ir$^{4+}$/Ir$^{5+}$ valency state where Ir$^{5+}$ should be non-magnetic in the strong SOC limit and to the perturbation of the Pr-O-Pr exchange pathways through the exchange of some Pr$^{3+}$ by Sr$^{2+}$.  The reduced value of magnetic entropy at $T_N$ in case of the Sr-doped compound compared to the undoped \PMIO~points to the persistence of short-range magnetic correlations. Our results reveal details of the interplay between, chemical doping, Ir-valence fluctuation, geometric frustration in shaping the magnetic ground states of  5$d$ iridate double perovskites.  Our work further provides valuable insights into the tunability of magnetic ground state behaviors in Ir-based complex double perovskite oxides and warrants future endeavors in designing candidate materials with exotic quantum magnetic ground states. 

\section{ACKNOWLEDGMENTS}
A.B. and D.T.A. thank the EPSRC UK for the funding (Grant No. EP/W00562X/1). D.D. thanks DST-INSPIRE for supporting fellowship. A.B. further acknowledges Lalit Narayan Mithila University for giving financial support. S.R. acknowledges the Science and Engineering Research Board (SERB), DST, India (Project no. CRG/2023/001060), and the UGC-DAE CSR scheme (CRS/2023-24/1646) for funding. A.B. and D.T.A. would like to thank the Royal Society of London for International Exchange funding between the UK and Japan, and Newton Advanced Fellowship funding between the UK and China. The authors acknowledge the TRC-DST of IACS, and the Materials Characterization Lab (MCL) of ISIS facility, UK, for providing experimental facilities and ILL for providing beam time. %\textcolor{red}{Exp No. xxxxx; Clemens: can you please put the number here?}.

\nocite{*}

\bibliography{sorsamp}% Produces the bibliography via BibTeX.

\end{document}